\documentclass[%
 reprint,
 amsmath,amssymb,
 aps,
]{revtex4-2}

\usepackage{graphicx}
\usepackage{dcolumn}
\usepackage{bm}

\begin{document}

\preprint{APS/123-QED}

\title{Formation mechanism of correspondence imaging with thermal light}

\author{Jian Leng,$^{1,2,\dag}$, Wen-Kai Yu,$^{1,2,\dag,}$}
\email{yuwenkai@bit.edu.cn}
\affiliation{$^1$Center for Quantum Technology Research, School of Physics, Beijing Institute of Technology, Beijing 100081, China\\
$^2$Key Laboratory of Advanced Optoelectronic Quantum Architecture and Measurements of Ministry of Education, School of Physics, Beijing Institute of Technology, Beijing 100081, China\\
$^\dag$These authors contributed equally to this work.\\}

\author{Shuo-Fei Wang,$^{1,2}$}
\affiliation{$^1$Center for Quantum Technology Research, School of Physics, Beijing Institute of Technology, Beijing 100081, China\\
$^2$Key Laboratory of Advanced Optoelectronic Quantum Architecture and Measurements of Ministry of Education, School of Physics, Beijing Institute of Technology, Beijing 100081, China\\
$^\dag$These authors contributed equally to this work.\\}



\date{\today}

\begin{abstract}
Correspondence imaging can achieve positive-negative ghost images by just conditional averaging of partial patterns, without treating bucket intensities as weights. To explain its imaging mechanism, we develop a probability theory assuming the targets are of gray-scale and the thermal reference speckles obey an arbitrary independent and identical distribution. By both simulation and experiments, we find that the recovered values in each region of the same original gray value conditionally obey a Gaussian distribution. A crosspoint-to-standard-deviation ratio is used as the figure of merit to prove that the patterns with respect to larger bucket values generate a positive image with a higher quality, vice versa for negative one. This work complements the theory of ghost imaging.
\end{abstract}

\maketitle


\section{\label{sec:level1}Introduction}
Ghost imaging (GI) provides a way to recover the object images via intensity correlation between reference patterns and bucket intensity signals. It was primitively demonstrated by using entangled light \cite{Pittman1995}, then was also experimentally realized with thermal or pseudo-thermal (a laser passing through a rotating ground glass) light \cite{DaZhang2005,Ferri2005,Jun2005,Liu2014}, as well as X-ray \cite{Hong2016,Zhang2018}. As long as the light field of reference arm and the object arm are conjugated, the lenses in GI with thermal light can be removed \cite{CaoPRA2005,Scarcelli2006}, which makes the imaging setup simpler and more flexible. Thus, the thermal light GI has been widely used in many fields, such as microscopic imaging \cite{YuOC2016}, optical encryption \cite{Clemente2010,YuAO2013,YuAO2019} and lidar \cite{Gong2015,YuSR2014}. To solve two key problems existing in GI, i.e., the image quality and measurement number, various GI methods have sprung up, such as background-removal GI \cite{Gatti2004}, differential GI (DGI) \cite{Ferri2010}, adaptive GI \cite{YuOE2014}, iterative denoising GI \cite{Yao2014}, blind GI \cite{Bertolotti2019}, super sub-Nyquist GI \cite{YuSensors2019}. Among these methods, the bucket values are served as the weights, reflecting the total intensities from the modulated object. Recently, an interesting experimental study found that one could generate the positive and negative ghost images by only conditionally averaging partial reference patterns. This method was named correspondence imaging (CI) \cite{Luo2011,Luo2012,Shih2012,MJSunAO2015}. It seemed that the bucket weights no longer participated in the correlation calculations involved in the second-order or high-order correlation functions, but actually they were completely binarized. Some confusing questions were raised, why could CI generate positive-negative images using only a few reference patterns, and why could CI work without involving bucket weights in the calculations? Their theoretical explanations have been the hot spots in this field for a long time, but after a few attempts \cite{Wen2012,YuCPB2015,YaoCOL2015}, researches were still exploring the path. Lately, a strict explantation based on a probability theory \cite{CaoPRA2018} was provided, which regarded the light intensities as stochastic variables and deduced a joint probability density function between the bucket and reference signals, giving us some inspiration. However, this theory was based on a fundamental assumption of the simplified model that consists of the negative exponential distributed light field and binary objects, thus it still had its limitations, especially in universality. The imaging mechanism of CI deserves further research.

In this paper, we assume a general model in which the targets are of gray-scale (each gray value has a large enough number of pixels), any two thermal speckles in the light field are independent of each other, all following an arbitrary identical distribution, and the whole reference speckles constitute a set of independent stochastic variables. The bucket values can be treated as many linear combinations of all pixels, also constituting a random variable. With above assumptions, we can deduce the joint probability density function between the bucket variable and each reference thermal speckle variable. After that, the forming formulas of the positive and negative images are also provided. Both simulation and experimental results have demonstrated the correctness of our derivation. Furthermore, we use this theoretical model to investigate how image quality varies with specific selection intervals used to average reference patterns.

\section{\label{sec:level2}Probability theory}
\subsection{\label{sec:level2.1}Statistical model of ghost imaging}
As we know, for a continuous random variable $X$, the probability of $X<x$ (i.e., the distribution function) can be written as $F_X(x)=P\{X<x\}$, then we have $F_X(-\infty)=0$, and $F_X(+\infty)=1$. Suppose the probability density function $f_X(x)$ of $X$ is the derivative of $F_X(x)$, i.e., $f_X(x)=F'_X(x)$, then
$\int_{-\infty}^{+\infty}f_X(x)dx=F_X(+\infty)-F_X(-\infty)=1$. Next, we will use its two typical mathematical properties of the random variable $X$, one is the mathematical expectation (also known as the mean) $E(X)$, defined as:
\begin{equation}
E(X)=\int_{-\infty}^{+\infty}xf_X(x)dx,
\end{equation}
the other is the variance $D(X)$, defined as:
\begin{align}
D(X)&=\int_{-\infty}^{+\infty}(x-E(x))^2f_X(x)dx\nonumber\\
	&=E(X^2)-E(X)^2.
\end{align}

We assume that the gray-scale object has a total of $M$ pixels, with $d$ representing the gray value of a pixel. The gray value of the $m$th point (pixel) is denoted by $d_m$, ranging from 0 to 1, with 0 being completely opaque and 1 being completely transparent. Accordingly, each reference pattern can also be divided into $M$ pixels, each of which has a light intensity expressed by $I_m$. This intensity value can be regarded as a random variable, which obeys an arbitrary identical probability distribution $I$. For simplicity of mathematics, it is assumed that the intensities of any two thermal speckles (pixels) in the reference spatial light field are statistically independent of each other. Then, the distribution function of the $m$th random variable $I_m$ can be written as $F_{I_m}(i_m)$, and its probability density function can be denoted by $f_{I_m}(i_m)$, where $i_m\in[0,\infty)$. The $m$th pixel of the object is illumined by the corresponding thermal speckle. On the plane after the thermal light passing through the gray-scale object, the $m$th point will have the value $d_mI_m$. Not only that, since there is still a certain distance between the object plane and the bucket detector, along with some existing influence factors such as diffraction, refraction, etc., a certain loss of light intensity should be considered here, expressed by the coefficient factor $a$. Thus, when the light reaches the sensing surface of bucket detector, the intensity becomes $Y_m=ad_mI_m$. Then, we have the following relationship between the $m$th point in the object arm and the $n$th point in the reference arm:
\begin{align}\label{eq:BE}
E(Y_mI_n)&=ad_mE(I_mI_n)\nonumber\\
	     &=\begin{cases}
		   ad_nE(I^2)& m=n,\\
		   ad_mE(I)^2& m\ne n.
           \end{cases}
\end{align}
The above formula is the the basic equation of ghost imaging with thermal light (GITL).

Besides, the bucket light intensity can be written as
\begin{equation}
S=\sum_{m}^M Y_m=a\sum_{m}^Md_mI_m,
\end{equation}
whose distribution function and probability density function are denoted by $F_S(s)$ and $f_S(s)(s\in[0,\infty))$, respectively.

For the convenience of calculation, suppose the subscript of the point of our interest is $n$, then we define a physical quantity $S_n$ that is very similar to the bucket value $S$, but excluding the bucket intensity with the subscript $n$:
\begin{equation}
S_n=\sum_{m\ne n}^M Y_m=a\sum_{m\ne n}^M d_mI_m.
\end{equation}
Obviously, $S_n$ is independent of $I_n$. According to the definition of $S_n$, we can immediately have
\begin{equation}
S=S_n+Y_n.
\end{equation}
We let $F_{S_n}(s_n)$ and $f_{S_n}(s_n)$ ($s_n\in[0,\infty)$) denote the distribution function and the probability density function of $S_n$, respectively.

With the above definitions, it is natural to calculate the second-order correlation $E(SI_n)$:
\begin{align}
E(SI_n)&=E[(S_n+Y_n)I_n]\nonumber\\
	   &=E(S_nI_n)+E(Y_nI_n)\nonumber\\
	   &=E\left[\left(\sum_{m\ne n}^M Y_m\right)I_n\right]+E(Y_nI_n)\nonumber\nonumber\\
	   &=\sum_{m\ne n}^ME(Y_mI_n)+E(Y_nI_n)\nonumber\\
	   &=\sum_{m\ne n}^Mad_mE(I)^2+ad_nE(I^2)\nonumber\\
	   &=a\sum_m^M d_mE(I)^2+a[E(I^2)-E(I)^2]d_n\nonumber\\
	   &=\gamma_1+\gamma_2d_n,
\end{align}
where both $\gamma_1$ and $\gamma_2$ are constants.

Since $d_n$ is the gray value of any object point, the physical meaning of the second-order correlation function is to perform the same linear transformation on the gray value of each object point. This is the essential reason why the second-order correlation algorithm can recover the object images. Thus, the basic formula of GITL, i.e., Eq.~(\ref{eq:BE}), plays a decisive role.

\subsection{\label{sec:level2.2}Approximation of model}
In this section, we begin by proving a theorem as follow to deduce the approximate distribution expressions of $S$ and $S_n$, which is only related to the mean $E(I)$ and the variance $D(I)$ of the light intensity $I$, independent of the specific distribution of $I$.

\noindent\textbf{Theorem 1:} \textit{When each gray value in the object image has infinite points (pixels), the bucket value} $S$ \textit{in GITL strictly obeys a normal distribution.}

\textit{Proof:} Let arbitrary gray value of the object be $d^{(k)}$ ($k\in\{1,2...K\}$), and its number of points (pixels) be $l^{(k)}$, which tends to infinity. We define the variable $S^{(k)}$ as the sum of all the points with the same gray value $d^{(k)}$ in the object arm as
\begin{align}
S^{(k)}&=\sum_{\{d_m=d^{(k)}\}}Y_m=\sum_{\{d_m=d^{(k)}\}}ad_mI_m\nonumber\\
	   &=ad^{(k)}\sum_{m=1}^{l^{(k)}}I_m.
\end{align}
Since $l^{(k)}$ tends to infinity, according to the central limit theorem for independently and identically distributed variables in the probability theory, $S^{(k)}$ follows a normal (Gaussian) distribution with a mean of $\mu^ {(k)}=l^{(k)}ad^{(k)}E(I)$ and a variance of $(\sigma^ {(k)})^2=l^{(k)}a^2(d^{(k)})^2D(I)$. Therefore, according to the gray value, we can rewrite the definition $S=\sum_m^M Y_m$ of $S$ as
\begin{equation}
S=\sum_m^M Y_m=\sum_k^K \left(\sum_{\{d_m=d^{(k)}\}}Y_m\right)=\sum_k^K S^{(k)}.
\end{equation}
Then, $S$ is the sum of $k$ Gaussian distributions. According to the probability theory, $S$ obeys a Gaussian distribution
\begin{align}
&F_S(s)\approx\phi\left(\frac{s-\mu}{\sigma}\right),\\
&f_S(s)\approx\frac{1}{\sqrt{2\pi}\sigma}e^{-\frac{(s-\mu)^2}{2\sigma^2}},
\end{align}
with a mean of
\begin{equation}
\mu=\sum_k^K \mu^{(k)}=\sum_k^K l^{(k)}ad^{(k)}E(I)=a\sum_m^Md_mE(I)
\end{equation}
and a variance of
\begin{align}
\sigma^2&=\sum_k^K (\sigma^{(k)})^2=\sum_k^K l^{(k)}a^2(d^{(k)})^2D(I)\nonumber\\
    	&=a^2\sum_m^Md_m^2D(I).\ \blacksquare
\end{align}

Thus, suppose each gray value owns sufficient points (pixels), the requirements of above theorem can be satisfied. Then, we will have that $S$ approximately follows a normal distribution with a mean of $\mu=a\sum_m^M d_mE(I)$ and a variance of $\sigma^2=a^2\sum_m^Md_m^2D(I)$. Similarly, $S_n$ also approximately fulfills a normal distribution with a mean of $\mu_n=a\sum_{m\ne n}^M d_mE(I)$ and a variance of $\sigma_n^2=a^2\sum_{m\ne n}^M d_m^2D(I)$.

\subsection{\label{sec:level2.3}Explaination for correspondence imaging}
With the obtained distributions of $S$ and $S_n$, we will start the calculation for CI. The joint probability density function between $S$ and $Y_n$ can be deduced as
\begin{align}
f_{S,Y_n}(s,y_n)&=f_{S_n}(s_n)\otimes f_{Y_n}(y_n)\nonumber\\
                &=f_{S_n}(s-y_n)f_{Y_n}(y_n).
\end{align}

To average the patterns corresponding to the bucket value $S$ above or below its ensemble average, we define
\begin{align}
&s_+=\begin{cases}
	 1&s\ge\mu,\\
	 0&s<\mu;
	 \end{cases}\\
&s_-=1-s_+.
\end{align}
Obviously, there are
\begin{align}
&\int s_+f_S(s)ds=\int_{\mu}^\infty f_S(s)ds=\frac{1}{2},\\
&\int s_-f_S(s)ds=\int (1-s_+)f_S(s)ds=\frac{1}{2}.
\end{align}

To obtain the average of the patterns that correspond to the bucket values above the ensemble average, i.e., $E(s_+I_n)$, we should first compute
\begin{align}
E(s_+Y_n)&=\frac{\int s_+y_nf_{S,Y_n}(s,y_n)dsdy_n}{\int s_+f_S(s)ds}\nonumber\\
         &=2\int_\mu^\infty\left[\int_0^sf_{S_n}(s-y_n)y_nf_{Y_n}(y_n)dy_n\right]ds.
\end{align}
Since $E(Y)\ll\mu$, we can treat $y_n$ in the above integral as a very small amount: $f_{S_n}(s-y_n)\approx f_{S_n}(s)-f'_{S_n}(s)y_n$. Besides, $s$ can be regarded as a very large amount: $\int_0^s\approx\int_0^\infty$. Then, we have
\begin{align}
       &E(s_+Y_n)\nonumber\\
\approx&2\int_\mu^\infty\left
\{\int_0^\infty[f_{S_n}(s)-f'_{S_n}(s)y_n]y_nf_{Y_n}(y_n)dy_n\right\}ds\nonumber\\
	  =&2E(Y_n)\int_\mu^\infty f_{S_n}(s)ds-2E(Y_n^2)\int_\mu^\infty f'_{S_n}(s)ds\nonumber\\
	  =&2E(Y_n)[1-F_{S_n}(\mu)]-2E(Y_n^2)[0-f_{S_n}(\mu)]\nonumber\\
	  =&2E(Y_n)\{1-F_{S_n}[\mu_n+E(Y_n)]\}\nonumber\\
       &+2E(Y_n^2)f_{S_n}[\mu_n+E(Y_n)]\nonumber\\
\approx&2E(Y_n)[1-F_{S_n}(\mu_n)-F'_{S_n}(\mu_n)E(Y_n)]\nonumber\\
       &+2E(Y_n^2)[f_{S_n}(\mu_n)+f'_{S_n}(\mu_n)E(Y_n)].
\end{align}
Since
\begin{align}
&F_{S_n}(\mu_n)=\frac{1}{2},\\
&F'_{S_n}(\mu_n)=f_{S_n}(\mu_n)=\frac{1}{\sqrt{2\pi}\sigma_n},\\
&f'_{S_n}(\mu_n)=0,
\end{align}
then
\begin{align}
E(s_+Y_n)\approx&2E(Y_n)[\frac{1}{2}-\frac{1}{\sqrt{2\pi}\sigma_n}E(Y_n)]+2E(Y_n^2)\frac{1}{\sqrt{2\pi}\sigma_n}\nonumber\\
               =&E(Y_n)+2D(Y_n)\frac{1}{\sqrt{2\pi}\sigma_n},
\end{align}
where
\begin{align}
&E(Y_n)=E(ad_nI_n)=ad_nE(I),\\
&D(Y_n)=D(ad_nI_n)=a^2d_n^2D(I),\\
&E(s_+Y_n)=E[s_+(ad_nI_n)]=ad_nE(s_+I_n).
\end{align}
So, we will get
\begin{equation}\label{eq:Ep}
E(s_+I_n)=E(I)+\sqrt{\frac{2}{\pi}}\frac{a}{\sigma_n}D(I)d_n.
\end{equation}
Use the standard deviation $\sigma=a\sqrt{\sum_m^Md_m^2D(I)}$ of $S$ to approximately replace the standard deviation $\sigma_n$ of $S_n$, we can acquire
\begin{align}
E(s_+I_n)&\approx E(I)+\sqrt{\frac{2D(I)}{\pi\sum_m^M d_m^2}}d_n\nonumber\\
         &=C_2+C_1d_n,
\end{align}
where
\begin{align}
&C_1=\sqrt{\frac{2D(I)}{\pi\sum_m^M d_m^2}}\\
&C_2=E(I).
\end{align}

Similarly, to calculate the average of the patterns that correspond to the bucket values below the ensemble average, i.e., $E(s_-I_n)$, we should first compute
\begin{align}
E(s_-Y_n)&=\frac{\int s_-y_nf_{S,Y_n}(s,y_n)dsdy_n}{\int s_-f_S(s)ds}\nonumber\\
         &=2\int(1-s_+)y_nf_{S,Y_n}(s,y_n)dsdy_n\nonumber\\
	     &=2E(Y)-E(s_+Y_n)\nonumber\\
	     &\approx E(Y_n)-2D(Y_n)\frac{1}{\sqrt{2\pi}\sigma_n}.
\end{align}
Using the exact same processing method as $E(s_+Y_n)$, we will have
\begin{align}
E(s_-I_n)&\approx E(I)-\sqrt{\frac{2D(I)}{\pi\sum_m^M d_m^2}}d_n\nonumber\\
	     &=C_2-C_1d_n.
\end{align}

Then we can compute the formula of the difference image:
\begin{align}
CI_\pm&=E(s_+I_n)-E(s_-I_n)\nonumber\\
      &=2C_1d_n.
\end{align}

Since $C_1$ and $C_2$ are all constants, the positive-negative images and $CI_\pm$ are all the linear transformations of the original object. For the reason that the efficient $C_1$ before $d_n$ in $E(s_+I_n)$ is positive, its result presents a positive image, while the efficient $-C_1$ before $d_n$ in $E(s_-I_n)$ is negative, the result is rendered as a negative image.

\section{\label{sec:level3}Verification for correspondence imaging}
The theoretical averages of the positive and negative images and $CI_\pm$ have been given above, but the gray value of each pixel in the actual reconstructed images generally fluctuates around the mean, following a certain distribution. Below, we will focus on this distribution and make a verification. Let us suppose there are a total of $T$ measurements, containing $T_+$ bucket values $S\ge\langle S\rangle$, and $T_-$ bucket values $S<\langle S\rangle$, where $\langle\cdots\rangle$ denotes the ensemble average of the signal. The operators $s_+$ and $s_-$ still use the definitions mentioned above. We denote the $n$th point in the $t$th speckle pattern as $I_{nt}$. Then, the positive and negative image formulas in the actual image reconstruction can be written as
\begin{align}
&\langle s_+I_n\rangle=\frac{1}{T_+}\sum_{t=1}^{T}s_{+}I_{nt},\\
&\langle s_-I_n\rangle=\frac{1}{T_-}\sum_{t=1}^{T}s_{-}I_{nt}.
\end{align}
According to the central limit theorem for independently and identically distributed variables, when $T_+$ is large enough, $\langle s_+I_n\rangle$ approximatively obeys a Gaussian distribution with a mean of $E(s_+I_n)$ and a variance of $\frac{D(s_+I_n)}{T_+}$; and similarly, when $T_-$ is large enough, $\langle s_-I_n\rangle$ approximatively follows a Gaussian distribution with a mean of $E(s_-I_n)$ and a variance of $\frac{D(s_-I_n)}{T_-}$.

Now, we will compute the variances $D(s_+I_n)$ and $D(s_-I_n)$. In a similar way of calculating $E(s_+I_n)$ and $E(s_-I_n)$, we first derive the following functions $E(s_+I_n^2)$ and $E(s_-I_n^2)$:
\begin{align}
&E(s_+I_n^2)\nonumber\\
\approx&E(I^2)+\sqrt{\frac{2}{\pi\sum_m^Md_m^2D(I)}}[E(I^3)-E(I^2)E(I)]d_n,\\
&E(s_-I_n^2)\nonumber\\
\approx&E(I^2)-\sqrt{\frac{2}{\pi\sum_m^Md_m^2D(I)}}[E(I^3)-E(I^2)E(I)]d_n.
\end{align}
By using the formula $D(X)=E(X^2)-E(X)^2$, the variances can be calculated as
\begin{align}
\label{eq:Dp}D(s_+I_n)&=E(s_+I_n^2)-E(s_+I_n)^2,\\
D(s_-I_n)&=E(s_-I_n^2)-E(s_-I_n)^2.
\end{align}

So far, we can theoretically calculate the distribution curve of a certain gray value $d^{(k)}$ (occupying a region that consists of several pixels) after reconstructing the images. For the positive image, it follows a Gaussian distribution with a mean of $E(s_+I_n)$ and a variance of $\frac{D(s_+I_n)}{T_+}$; while for the negative image, it obeys a Gaussian distribution with a mean of $E(s_-I_n)$ and a variance of $\frac{D(s_-I_n)}{T_-}$. In both simulation and experiments, we calculate the probability of the recovered pixel values falling in each pixel region where the gray value of the original image equals $d^{(k)}$, and plot the corresponding probability density curves, compared with the theoretical Gaussian curve to demonstrate the correctness of the theory. The Gaussian distribution theoretical curves are obtained from the computed theoretical means and variances.

\subsection{\label{sec:level3.1}Simulation}
\begin{figure}[htbp]
\centering
\includegraphics[width=0.9\linewidth]{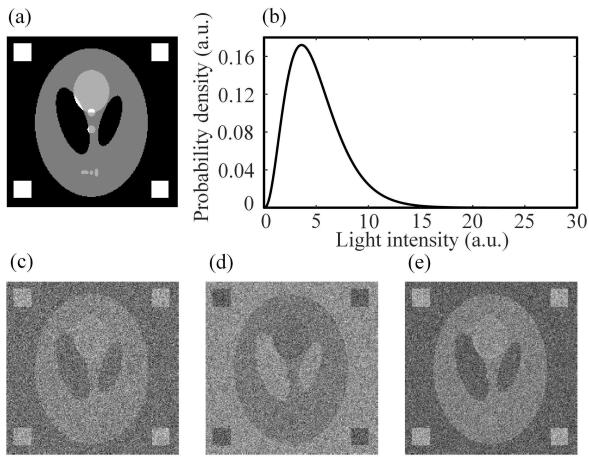}
\caption{\label{fig:simulation}Simulation results. (a) is the original image, a modified head phantom image. (b) is the chosen probability density function curve. (c)--(e) are the reconstructions of positive-negative images and their difference image, respectively.}
\end{figure}

Here, we chose an object image of $200\times200$ pixels, as shown in Fig.~\ref{fig:simulation}(a), and its statistical data of the gray values was given in Table~\ref{tab:table1}. We took the speckle variables of the patterns which obeyed an identical gamma distribution for an example, and the gamma distribution was parameterized in terms of a shape parameter $\alpha=3.57$ and the a scale parameter $\theta=1.4$, and its probability density function could be expressed as
\begin{equation}
f_I(i)=\frac{i^{\alpha-1}e^{-i/\theta}}{\theta^\alpha\Gamma(\alpha)},\textrm{ for }i>0,
\end{equation}
as plotted in Fig.~\ref{fig:simulation}(b). The positive and negative images with a total of 50000 frames and their difference image $CI_\pm$ were given in Figs.~\ref{fig:simulation}(c)--\ref{fig:simulation}(e).
\begin{table}[htbp]
\caption{\label{tab:table1}Statistical data of gray values in the original image}
\begin{ruledtabular}
\begin{tabular}{ccc}
Gray value&Total number of pixels&Proportion\\
\colrule
0 & 23353 & 58.38\%\\
0.5 & 13147 & 32.87\%\\
0.7 & 1733 & 4.33\%\\
1 & 1767 & 4.42\%\\
\end{tabular}
\end{ruledtabular}
\end{table}

\begin{figure}[htbp]
\centering
\includegraphics[width=0.9\linewidth]{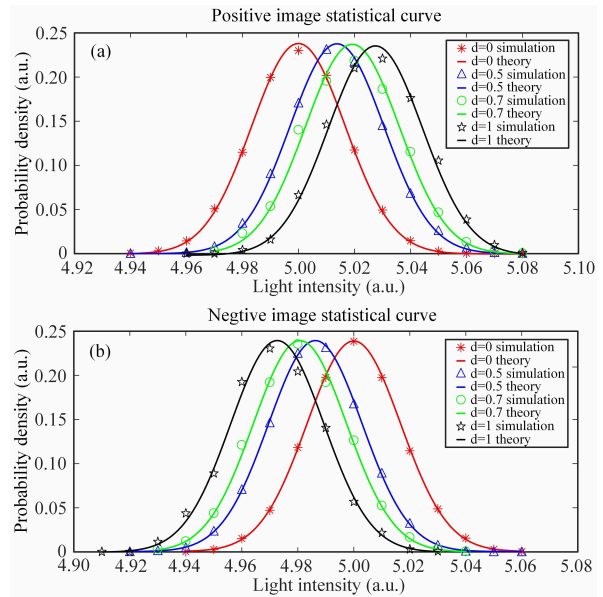}
\caption{\label{fig:PDF}Probability density function curves for the recovered pixel values, compared with the theoretical gamma function curves. (a)--(b) are the probability density distributions and theoretical gamma curves of reconstructed pixel values falling in each pixel region where the gray value of the original image equals $d^{(k)}$, for positive and negative images, respectively. The abscissa is the reconstructed pixel value, and the ordinate indicates the probability of occurrence of these values.}
\end{figure}

Then, for both positive and negative images, we separately computed the probability of the reconstructed pixel values falling in each pixel region corresponding the one that consists of pixel positions with the same gray value $d^{(k)}$ of the original image, and drew their probability density curves to compare with the gamma theoretical curves, as shown in Figs.~\ref{fig:PDF}(a)--\ref{fig:PDF}(b). From the graphs, we could clearly see that the recovered pixel value data is highly consistent with the presupposed gamma distribution.

\subsection{\label{sec:level3.2}Experiment}
For the practical optical experiments, there are many kinds of noise. it is hard to determine the noise distribution, but the superposition of multiple probability distributions will result in a Gaussian distribution with a large probability. In this case, we may assume the measurement noise fulfills Gaussian statistics. In a similar way, suppose the Gaussian noise is a random variable, denoted by $X$, with a mean of $E(X)=0$ and an unknown variance $D(X)$. We add this noise to the bucket variable, then get
\begin{equation}
S=S_n+Y_n+X.
\end{equation}
The same as the previous discussion, one only needs to replace the previous $S_n$ with $S_n+X$ for the calculation in actual measurement environment. And $S_n+X$ satisfies a Gaussian distribution with a mean $\mu_n+E(X)$ and a variance $\sigma_n^2+D(X)$. Here, we directly present the results:
\begin{align}
E(s_+I_n)\approx& E(I)+\sqrt{\frac{2}{\pi}}\sqrt{\frac{1}{\sum_m^M d_m^2D(I)+\frac{D(X)}{a^2}}}D(I)d_n,\\
E(s_-I_n)\approx& E(I)-\sqrt{\frac{2}{\pi}}\sqrt{\frac{1}{\sum_m^M d_m^2D(I)+\frac{D(X)}{a^2}}}D(I)d_n,\\
E(s_+I_n^2)\approx& E(I^2)+\sqrt{\frac{2}{\pi}}\sqrt{\frac{1}{\sum_m^Md_m^2D(I)+\frac{D(X)}{a^2}}}\nonumber\\
&\times[E(I^3)-E(I^2)E(I)]d_n,\\
E(s_-I_n^2)\approx&E(I^2)-\sqrt{\frac{2}{\pi}}\sqrt{\frac{1}{\sum_m^Md_m^2D(I)+\frac{D(X)}{a^2}}}\nonumber\\
&\times[E(I^3)-E(I^2)E(I)]d_n,\\
D(s_+I_n)=&E(s_+I_n^2)-E(s_+I_n)^2,\\
D(s_-I_n)=&E(s_-I_n^2)-E(s_-I_n)^2.
\end{align}
There is only one pending term introduced by noise and light intensity attenuation, i.e., $\frac{D(X)}{a^2}$. It is hard for us to know its specific value. This can only be obtained empirically in order to match the experimental data to the theoretical curve as much as possible.

\begin{figure}[htbp]
\centering
\includegraphics[width=0.95\linewidth]{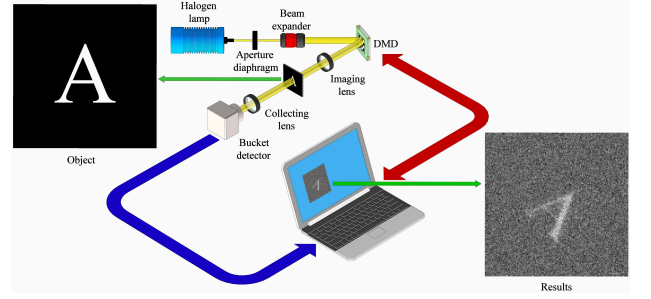}
\caption{\label{fig:setup}Optical setup for CI. The thermal light emitted from a halogen lamp passes through an aperture diaphragm and a beam expander, and illuminates a DMD. Then, the modulated light is projected onto a black-and-white film (i.e., the object). The total intensities are recorded by a bucket detector.}
\end{figure}

\begin{figure}[htbp]
\centering
\includegraphics[width=0.9\linewidth]{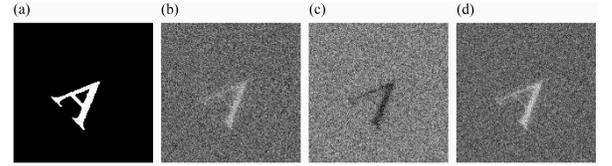}
\caption{\label{fig:expresults}Experimental results. (a) is the binarized image obtained by a camera. (b)--(d) are the recovered positive-negative images and their difference image, respectively.}
\end{figure}

\begin{figure}[htbp]
\centering
\includegraphics[width=0.9\linewidth]{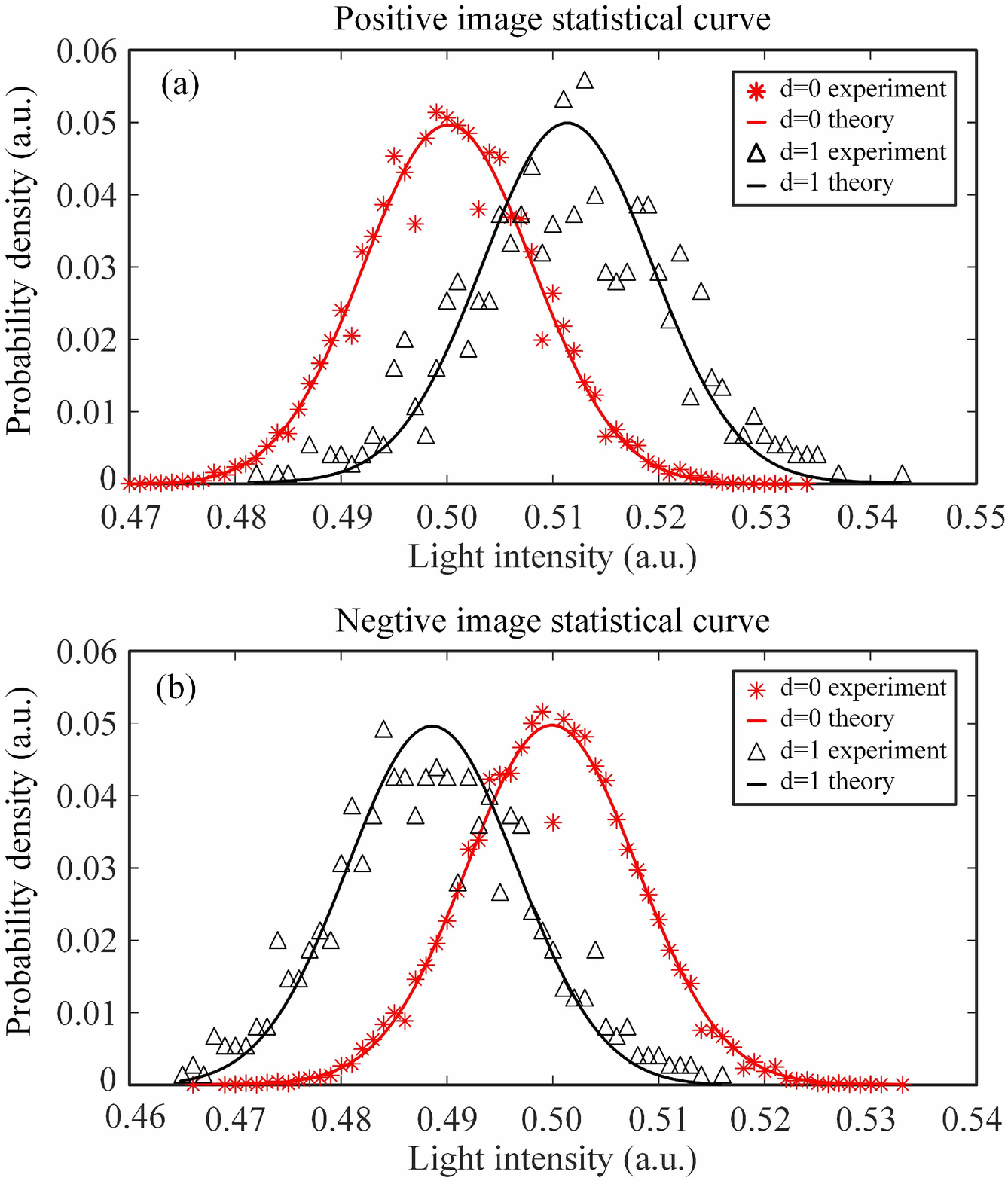}
\caption{\label{fig:expPDF}Probability density function curves for the recovered pixel values, compared with theoretical Gaussian function curves. (a)--(b) are the probability density distributions and theoretical Gaussian curves of recovered pixel values falling in each pixel region where the original gray value equals 0 or 1, for positive and negative images, respectively. Here, the value of the pending term $\frac{D(X)}{a^2}$ is set to 120.}
\end{figure}

Our experiment was based on a widely used computational GI setup, as shown in Fig.~\ref{fig:setup}. Unlike double arm GI, it could modulate the illumination light according to the preset patterns without the help of an array detector with spatial resolution. A digital micromirror device (DMD) which consisted of 1,024 $\times$ 768 micro-mirrors, each of size $13.68\times13.68\ \mu\textrm{m}^2$, was used here to perform light intensity modulation. Since each of its micromirror could be oriented either +12$^\circ$ and -12$^\circ$ with respect to the normal of the DMD work plane, corresponding to the bright pixel 1 or the dark pixel 0, the light would be reflected into two directions. In our experiment, the light from a halogen lamp illuminated the DMD through an aperture diaphragm and a beam expander, then the modulated patterns were projected onto an object, which was a black-and-white film printed with ``A'', as shown in Fig.~\ref{fig:expresults}(a). Its statistical data of binary values was provided in Table~\ref{tab:table2}. The 0-1 random patterns used occupied the central $160\times160$ micromirrors (pixels) of the DMD. In each pattern, 0 and 1 had the same probability of occurrence. A 1/1.8 inch charge-coupled device (CCD) was used as a bucket detector to integrate the gray values of all pixels in one frame. The recovered images with 7761 frames were presented in Figs.~\ref{fig:expresults}(b)--\ref{fig:expresults}(d). From the curves shown in Fig.~\ref{fig:expPDF}, the experimental data was in good agreement with the theoretical Gaussian curves.
\begin{table}[htbp]
\caption{\label{tab:table2}Statistical data of binary values in the binarized image taken by a camera}
\begin{ruledtabular}
\begin{tabular}{ccc}
Gray value&Total number of pixels&Proportion\\
\colrule
0 & 24847 & 97.06\%\\
1 & 753 & 2.94\%\\
\end{tabular}
\end{ruledtabular}
\end{table}

\section{\label{sec:level4}Conditional-averaging ghost imaging with a potential application}
\begin{figure}[htbp]
\centering
\includegraphics[width=0.9\linewidth]{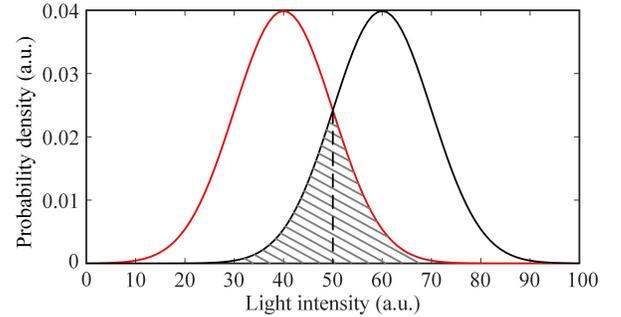}
\caption{\label{fig:twocurves}Schematic diagram of two Gaussian curves corresponding to two gray values.}
\end{figure}

As mentioned before, the statistical curve of each gray value within a certain pixel region in the positive or negative image corresponds to a Gaussian curve. In Fig.~\ref{fig:twocurves}, we drew two Gaussian curves obtained from two pixel regions corresponding to two gray values. Obviously, the farther the Gaussian curves of two gray values are separated, the bigger is the difference between the two recovered gray values, and the better is the image quality of the reconstructed. We can choose an appropriate measure to describe this distance, e.g., the overlapping area of two curves, denoted by $\Omega$, which can be treated as a criterion for the reconstruction quality.

Analogously, it is easy to find that for the reconstructed images using the correlation functions, such as $G_2=\langle S\cdot I_n\rangle$, $g_2=\frac{\langle S\cdot I_n\rangle}{\langle S\rangle\langle I_n\rangle}$, $DGI=\langle S\cdot I_n\rangle-\frac{\langle S\rangle}{\langle S_R\rangle}\langle S_R\cdot I_n\rangle$, etc., the conclusion that the reconstructed pixels in each pixel region obey a Gaussian or Gaussian-like distribution is still valid. Thereby, these functions can also use this overlapping area as the image quality measure.

Now, let us calculate this overlapping area $\Omega$. In Fig.~\ref{fig:twocurves}, the two curves that correspond to any two original gray values $\varsigma$ and $\tau$ have two means, i.e., $\mu_1$ and $\mu_2$. Generally, as long as the algorithm can reconstruct the object image, it is obvious that there must be a linear relationship between the reconstructed image and the original image, which will be at most affected by noise. For simplicity of mathematics, we suppose the standard deviations are approximately equal, i.e., $\sigma_1\approx\sigma_2=\sigma$. Actually, in both simulation and experiments, we also observed that the standard deviations of the Gaussian curves for all different original gray values were very close to each other. Because the original speckle intensities are independent and identically distributed, when the number of pixels contained in each pixel region is large enough, the standard deviations of the average values of the reference patterns inside these pixel regions will also tend to the same value. Without loss of generality, we can set $\mu_1<\mu_2$. It is easy to calculate the abscissa of the intersection of two curves, i.e., $\frac{\mu_1+\mu_2}{2}$. The shaded area in Fig.~\ref{fig:twocurves} is $\Omega=2\phi(-\frac{\mu_2-\mu_1}{2\sigma})$, where $\phi(x)$ is the standard Gaussian distribution function (the integral of the standard Gaussian probability density function). Note that the area is negatively correlated with $\frac{\mu_2-\mu_1}{2\sigma}$, which is a term related to the original gray values $\varsigma$ and $\tau$. If the standard deviations are assumed to be approximately equal, then the well-known formula of contrast-to-noise ratio (CNR) \cite{Chan2010} differs from this term $\frac{\mu_2-\mu_1}{2\sigma}$ only by a constant factor $\sqrt{2}$. To some extent, for binary objects, the CNR is a special case of the overlapping area, and can be derived from the latter, thus the physical meaning of CNR is manifested here. However, $\frac{\mu_2-\mu_1}{2\sigma}$ is not very suitable as an assessment metric of reconstruction quality for the following reasons. For the same reconstruction image, the value of $\frac{\mu_2-\mu_1}{2\sigma}$ calculated from two distant original gray-scale values (such as 0 and 1) is much larger than that of two original gray values which are close to each other (e.g., 0.4 and 0.6), but it does not mean that the former result is much better than the latter. Because they are all obtained from the same recovered image, the former get a larger value since they are calculated from two original gray-scale values that are much easier to be resolved. To provide a fair comparison, we will introduce a new imaging quality factor named crosspoint-to-standard-deviation ratio (CSR), which is defined as
\begin{equation}
\textrm{CSR}=\frac{(\mu_2-\mu_1)/2}{\sigma}\delta,
\end{equation}
where $\delta=\frac{1}{\tau-\varsigma}$. Since an identical linear relationship is associated with the original gray values and the means, the product between the terms $\frac{1}{\tau-\varsigma}$ and $\mu_2-\mu_1$ eliminates the effects of the specific gray values so that the CSR values obtained by choosing any two original gray values for the reconstructed images are the same. For any two given original gray values, the larger is the CSR value, the smaller is the overlapping area, and the more obviously the two gray values are separated, the better is the imaging quality.

As mentioned above, the positive or negative image are obtained by just averaging partial reference patterns corresponding to the bucket values above or below a threshold. Now, we will use the CSR to discuss the effect of using different intervals of partial reference patterns on the reconstruction quality. For the positive image, we define a logic signal,
\begin{equation}
s_\beta=\begin{cases}
		   1&s\ge\beta\mu,\\
		   0&s<\beta\mu.
		   \end{cases}
\end{equation}

The number of the patterns that correspond to the bucket values larger than $\beta\mu$ is $T_\beta=T[\int_{\beta\mu}^\infty f_S(s)ds]=T[1-F_S(\beta\mu)]$, where $T$ is the total number of measurements. Then, we can acquire
\begin{align}
&E(s_\beta Y_n)\nonumber\\
=&\frac{\int s_\beta y_nf_{S,Y_n}(s,y_n)dsdy_n}{\int s_\beta f_S(s)ds}\nonumber\\
=&\frac{\int_{\beta \mu}^\infty\left[\int_0^s f_{S_n}(s-y_n)y_nf_{Y_n}(y_n)dy_n\right]ds}{\int_{\beta\mu}^\infty f_S(s)ds}\nonumber\\
\approx&\frac{\int_{\beta\mu}^\infty\left\{\int_0^\infty [f_{S_n}(s)-f'_{S_n}(s)y_n]y_nf_{Y_n}(y_n)dy_n\right\}ds}{1-F_S(\beta\mu)}\nonumber\\ =&\frac{E(Y_n)[1-F_{S_n}(\beta\mu)]-E(Y_n^2)[0-f_{S_n}(\beta\mu)]}{1-F_S(\beta\mu)}\nonumber\\
=&\frac{E(Y_n)[1-F_{S_n}(\beta\mu)]+E(Y_n^2)f_{S_n}(\beta\mu)}{1-F_S(\beta\mu)}.
\end{align}
In a similar way, we acquire the formula of $E(s_\beta Y_n^2)$:
\begin{equation}
E(s_\beta Y_n^2)=\frac{E(Y_n^2)[1-F_{S_n}(\beta\mu)]+E(Y_n^3)f_{S_n}(\beta\mu)}{1-F_S(\beta\mu)}.
\end{equation}
Then, there are
\begin{align}
E(s_\beta I_n)&=\frac{E(I)[1-F_{S_n}(\beta\mu)]+aE(I^2)f_{S_n}(\beta\mu)d_n}{1-F_S(\beta\mu)},\\
E(s_\beta I_n^2)&=\frac{E(I^2)[1-F_{S_n}(\beta\mu)]+aE(I^3)f_{S_n}(\beta\mu)d_n}{1-F_S(\beta\mu)}.
\end{align}
Thus, the CSR can be written as
\begin{equation}\label{eq:CSR}
\textrm{CSR}=\frac{|E(s_\beta I_n)|_{d_n=\varsigma}-E(s_\beta I_n)|_{d_n=\tau}|}{2\sqrt{\frac{|E(s_\beta I_n^2)|_{d_n=\tau}-E(s_\beta I_n)^2|_{d_n=\tau}|}{T_\beta}}}\frac{1}{|\varsigma-\tau|}.
\end{equation}

Now, we will discuss the generality of CSR to obtain the trend of CSR changing with $\beta$ without pursuing its specific values. In Eq.~(\ref{eq:CSR}), since each gray value has little effect on the standard deviation, we set $\tau$ in the denominator equals 0; because $E(I)\ll\mu$, the distributions of $S$ and $S_n$ can be considered to be approximately the same, and $\beta\mu$ is not much different from $\mu$. Then, the CSR formula can be simplified to
\begin{equation}
\textrm{CSR}=\frac{aE(I^2)f_S(\beta\mu)T}{2[1-F_S(\beta\mu)]^{\frac{1}{2}}\sqrt{D(I)}}.
\end{equation}
Obviously, the larger is the total number of measurements, the higher is the CSR value, and the better is the reconstruction quality. Apart from this, the CSR value also depends on the following factor
\begin{equation}
g(\beta\mu)=\frac{f_S(\beta\mu)}{[1-F_S(\beta\mu)]^{\frac{1}{2}}}.
\end{equation}
We take the derivative of this factor with respect to $\beta\mu$ (there is $f^\prime_S(\beta\mu)=0$ under first-order approximation):
\begin{equation}
g^\prime(\beta\mu)=\frac{\frac{1}{2}f_S^2(\beta\mu)}{[1-F_S(\beta\mu)]^\frac{3}{2}}>0.
\end{equation}
It can be concluded that $g(\beta\mu)$ is an increasing function, and the CSR value increases gradually as $\beta$ increases. It means that the patterns that correspond to much larger bucket values (above the mean) will undoubtedly generate a positive image with much higher quality, and vice versa for the negative image formation. It also helps to explain the inner mechanism of the previous work in super sub-Nyquist single-pixel imaging \cite{YuSensors2019}.

\section{\label{sec:level5}Conclution}
In summary, we have developed a probability theory to explain the formation mechanism of CI whose the bucket values are binarized, based on a general model in which the targets are of gray-scale, and any two thermal reference speckles are independent of each other, all following an arbitrary identical distribution. By building the joint probability density function between the bucket variable and each reference thermal speckle variable, and deducing the related means and variances, we find that the positive-negative images and their difference image are all the linear transformations of the object image. Provided that each original gray-scale value has a large enough number of pixels, then the reconstructed values falling in every pixel region of the same original gray value will obey a Gaussian distribution, no matter what kind of distribution the speckles obey. The measurement noise is also considered. We have demonstrated the validity of the derived formulas through both simulation and experiments. On the basis of our theory, we also introduce a new image quality metric CSR, and prove that the patterns that correspond to much larger bucket values (above the mean) will help generate a positive image of much higher quality, and vice versa for the negative one. Therefore, this work will give rise to many potential practical applications.

\begin{acknowledgments}
This work was supported by the Natural Science Foundation of Beijing Municipality (Grant No. 4184098), the National Natural Science Foundation of China (Grant No. 61801022), the National Key Research and Development Program of China (Grant No. 2016YFE0131500), the Civil Space Project of China (Grant No. D040301), the International Science and Technology Cooperation Special Project of Beijing Institute of Technology (Grant No. GZ2018185101), and the Beijing Excellent Talents Cultivation Project - Youth Backbone Individual Project.
\end{acknowledgments}

\nocite{*}


\end{document}